\documentclass[prb,amsmath,amssymb,showpacs,1superscriptaddress]{revtex4}

\usepackage{graphicx,psfrag,epsfig}

\begin{document}
\title{Electron injection in a nanotube with leads: finite\\
frequency noise-correlations and anomalous charges}
%\title{Finite frequency noise cross-correlations in a\\
%nanotube connected to Fermi liquid leads}

\author{Andrei V. Lebedev$^{a,b}$}
\author{Adeline Cr\'epieux$^b$}
\author{Thierry Martin$^b$}

\affiliation{$^a$Landau Institute for Theoretical Physics RAS, 117940 Moscow, Russia\\
$^b$Centre de Physique Th\'eorique, Universit\'e de la
M\'editerran\'ee, Case 907, 13288 Marseille, France}

\begin{abstract}
The non-equilibrium transport properties of a carbon nanotube which is connected to Fermi liquid 
leads, where electrons are injected in the bulk, are computed. A previous work which considered 
an infinite nanotube showed that the zero frequency noise correlations, measured at opposite ends 
of the nanotube, could be used to extract the anomalous charges of the chiral excitations which 
propagate in the nanotube. Here, the presence of the leads have the effect that such-noise 
cross-correlations vanish at zero frequency. Nevertheless, information concerning the 
anomalous charges can be recovered when considering the spectral density of noise correlations 
at finite frequencies, which is computed perturbatively in the tunneling amplitude.
The spectrum of the noise cross-correlations is shown to depend crucially on the ratio of the 
time of flight of quasiparticles traveling in the nanotube to the ``voltage'' time which 
defines the width of the quasiparticle wave-packets injected when an electron tunnels. 
Potential applications toward the measurement of such anomalous charges in non-chiral 
Luttinger liquids (nanotubes or semiconductor quantum wires) are discussed.   
\end{abstract}
\pacs{}

\maketitle

\section{Introduction}

Over the years nano-electronics has been dealing mostly with transport through
artificial nano-structures, made from semiconductor or from metallic material.
Nowadays, attention has also focused on individual nano objects
connected to leads: molecules, conjugated polymers, carbon nanotubes
among others. Carbon nanotubes are especially interesting because,
depending on their helicity, they constitute a near ideal one-dimensional
metal\cite{nanotube}. Indeed, the dispersion of metallic nanotubes at
the Fermi level is linear, and one can therefore
expect that electron correlations play a dominant role\cite{egger_review}.
Evidence for Luttinger liquid behavior has been seen in tunneling experiments\cite{mceuen}.
A non-linear current-voltage characteristic\cite{kane_balents_fisher}
was predicted for both tunneling
in the bulk of the nanotube and to the end of a nanotube. The tunneling density
of states exponent found in these experiments suggest that the Luttinger
liquid interaction parameter describing the total charge modes $K_{c+}\simeq 0.3$, which should be compared to the
non-interacting value  $K_{c+}=1$: the effect of interactions is strong.
A nanotube therefore ``does not like'' to accommodate electrons, because its
elementary excitations do not resemble electrons as in a Fermi Liquid (FL):
they consist of the collective bosonic excitations which are known to occur
in one-dimensional correlated electron systems \cite{haldane}.

In chiral Luttinger liquids, which are believed to describe the physics of the
fractional quantum Hall edge states, a measurement of the backscattering noise
is sufficient to identify the fractional charge \cite{fqhe} of the quasiparticle.
No straightforward analogy exists for a quantum wire or for a nanotube.
Recent work have shown\cite{ines_annales,pham}
that although left and right moving electrons mix in such Non-Chiral Luttinger
Liquids (NCLL), their elementary excitations can be decomposed in right and left
moving chiral bosonic modes which carry a non integer electron charge,
where the latter depends on the interaction parameter identified below as $K_{c+}$.

\begin{figure}[h]
\epsfxsize 7 cm
\centerline{\epsffile{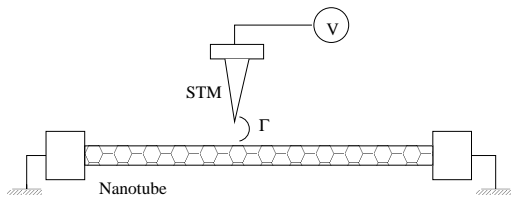}}
\caption{Nanotube connected to leads, with electrons injected in the bulk via a STM
\label{fig1}}
\end{figure}

\begin{figure}[h]
\epsfxsize 7 cm
\centerline{\epsffile{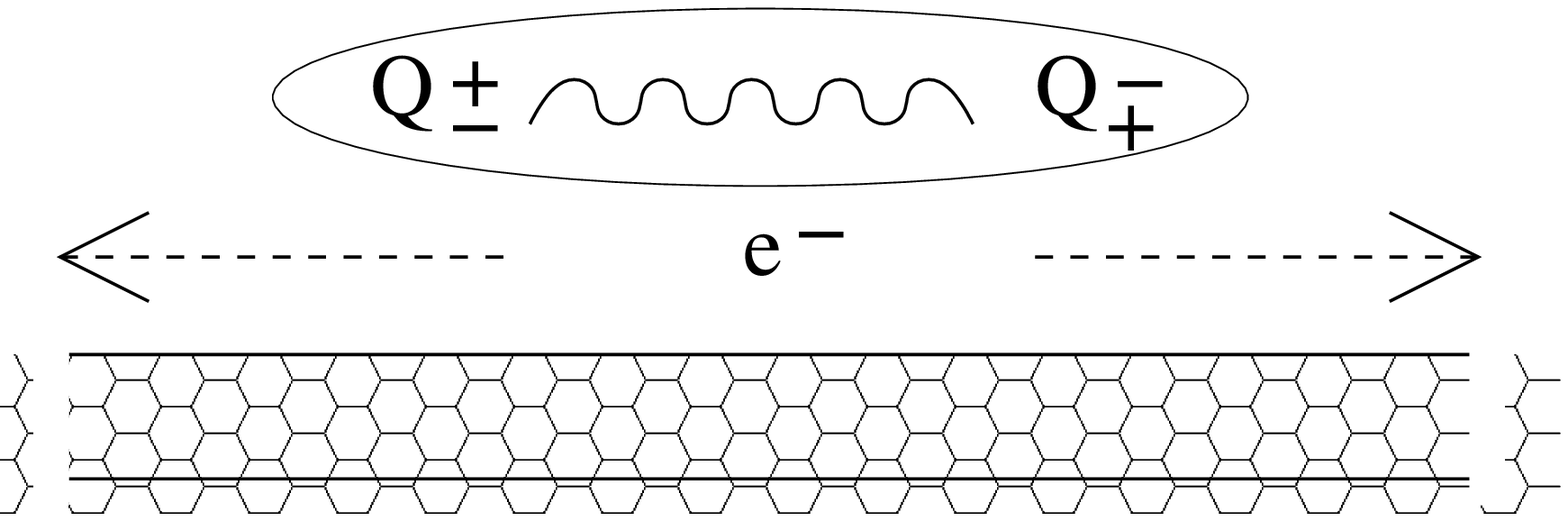}}
\caption{Entanglement of quasiparticle excitations in an infinite Luttinger liquid.
\label{fig2}}
\end{figure}

Recently a theoretical suggestion to measure these anomalous charges using
both current noise auto and cross-correlations was proposed\cite{crepieux}.
Contrary to the fractional quantum Hall effect, an autocorrelation noise
measurement alone is not sufficient to isolate the anomalous charge of a NCLL.
In Ref. \onlinecite{crepieux}, it was assumed that electrons were injected
into the bulk of the nanotube (Fig. \ref{fig1}), while current was being measured at both
ends. The electron is then split into two chiral quasi-particle modes which move
in opposite directions. Each pair of modes carries either of two anomalous
charges $Q_\pm=(1\pm K_{c+})/2$ attached to the right/left movers.
Yet because an electron is injected locally,
it has an equal probability to have $Q_+$($Q_-$) on the right or on the
left (Fig. \ref{fig2}): the wave function describing the injection of an electron has
entangled quasi-particle degree of freedom, with quantum numbers,
or states, specified by the anomalous charges $Q_\pm$.

\begin{figure}[h]
\epsfxsize 7 cm
\centerline{\epsffile{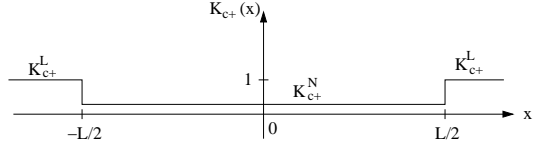}}
\caption{1D model for non interacting leads: the Luttinger liquid parameter
varies over distance.\label{fig3}}
\end{figure}

Nevertheless, one drawback is that injection was studied in an infinite
nanotube. In the last ten years, it has been argued that when a Luttinger
liquid is connected to normal metal leads, one looses the possibility
for detecting anomalous charges in NCLL. One of such models consists of
a Luttinger liquid with an interaction parameter which varies as a
function of distance:  $K_{c+}=1$ in the leads while $K_{c+}<1$ in the Luttinger liquid
(Fig. \ref{fig3}). The purpose of the present work is threefold: first, the results
for an infinite nanotube at zero frequencies will be extended to the finite frequency domain.
Second, it will be
shown explicitly that the zero frequency noise cross correlations vanish when
one includes the non interacting leads described above. This result thus
suggests that anomalous charges cannot be detected by the above correlation method.
Third, are the good news: an analysis of finite frequency autocorrelation noise
and noise cross-correlations -- in the presence of Fermi liquid leads -- allows to
recover crucial information about the anomalous charges.
The time of flight $\tau_\mathrm{L}=L/2v_{c+}$ for excitations in the nanotube
and the voltage time scale $\tau_\mathrm{V}=\hbar/eV$ are the sole parameters which
specify the behavior of the noise cross-correlations as a function
of frequency.

\section{Model Hamiltonian}

Consider the following experimental setup: a nanotube connected
to a two FL leads. An STM tip is put in contact with
the bulk nanotube at $x=0$. A bias voltage $V$ is applied
between the STM tip and the nanotube allowing electron transfer. Subsequently,
quasi-particle excitations propagate along the nanotube
towards the FL leads. Following
Ref.~\onlinecite{egger_98}, the electron field
$\Psi_{r\alpha\sigma}(x,t)$ in the nanotube, describing the
electron moving along the direction $r=\pm 1$, from the mode
$\alpha=\pm 1$ with a spin $\sigma=\pm 1$, can be written in terms
of the bosonic field $\varphi_{r\alpha\sigma}(x,t)$ as
\begin{equation}
      \Psi_{r\alpha\sigma}(x,t)=\frac{F_{r\alpha\sigma}}{
      \sqrt{2\pi a}}\,e^{ik_{\rm\scriptscriptstyle F}rx+
      iq_{\rm\scriptscriptstyle F}\alpha x+i\varphi_{r\alpha\sigma}
      (x,t)},
\end{equation}
where $a$ is an ultraviolet cutoff of the Luttinger Liquid (LL)
model, $F_{r\alpha\sigma}$ are the Klein factors,
$k_{\scriptscriptstyle\rm F}$ the Fermi momentum and
$q_{\scriptscriptstyle\rm F}$ is the momentum mismatch associated
with the two modes. For further calculation purposes it is
convenient to rewrite the bosonic field
$\varphi_{r\alpha\sigma}(x,t)$ in terms of the non-chiral bosonic
fields $\theta_{j\delta}$ and $\phi_{j\delta}$, with index
$j\delta\in\{c+,c-,s+,s-\}$ identifying the charge/spin,
total/relative fields:
\begin{equation}
      \varphi_{r\alpha\sigma}(x,t)=\frac{\sqrt{\pi}}2
      \sum\limits_{j\delta}h_{\alpha\sigma j\delta}\,
      \bigl[\phi_{j\delta}(x,t)+r\theta_{j\delta}(x,t)
      \bigr],
\end{equation}
with factors $h_{\alpha\sigma c+}=1$, $h_{\alpha\sigma
c-}=\alpha$, $h_{\alpha\sigma s+}=\sigma$ and $h_{\alpha\sigma
s-}=\alpha\sigma$, and bosonic fields obeying the equal time
commutation relations
$[\phi_{j\delta}(x),\theta_{j^\prime\delta^\prime}(x^\prime)]=-(i/2)
\delta_{jj^\prime}\delta_{\delta\delta^\prime}\,sgn(x-x^\prime)$.
The Hamiltonian describing the nanotube connected to FL leads
assumes the form:
\begin{eqnarray}
      H&=&\frac12\sum\limits_{j\delta}\int\limits_{-\infty}^\infty
      dx\;v_{j\delta}(x)\Bigl[K_{j\delta}(x)(\partial_x\phi_{j\delta})^2
      \nonumber\\
      &\ &\qquad+
      K_{j\delta}^{-1}(x)(\partial_x\theta_{j\delta})^2\Bigr],
      \label{b_ham}
\end{eqnarray}
where the interaction parameters $K_{j\delta}(x)$ is now
assumed to depend on position and the velocities $v_{j\delta}(x)$ satisfy
$v_{j\delta}(x)=v_{\scriptscriptstyle\rm
F}/K_{j\delta}(x)$.

The electrons in STM tip are assumed to be non-interacting.
For convenience\cite{crepieux}, the electron field $c_\sigma(t)$ in the STM
tip can be described in terms of a semi-infinite LL with interaction parameter equal to 1:
\begin{equation}
      c_\sigma(t)=\frac{f_\sigma}{\sqrt{2\pi a}}\,e^{i\tilde
      \varphi_\sigma(t)},
\end{equation}
where $\tilde \varphi_\sigma(t)$ is the chiral bosonic field,
whose Keldysh Green's function at $x=0$ is given
by~\cite{chamon_99}:
\begin{eqnarray}
      &&g_{\eta\mu}^\sigma(t_1-t_2)=\Bigl\langle
      T_\mathrm{K}\bigl\{\tilde\varphi_{\sigma}(t_1^{\eta})
      \tilde\varphi_{\sigma}(t_2^{\mu})\bigr\}\Bigr\rangle
      \label{STM_gr}\\
      &&=\!-\!\ln\!\left[1+i(\eta\!+\!\mu)
      \frac{v_{\scriptscriptstyle\rm F}
      |t_1\!-\!t_2|}{2a}-i(\eta\!-\!\mu)
      \frac{v_{\scriptscriptstyle\rm F}
      (t_1\!-\!t_2)}{2a}\right],\nonumber
\end{eqnarray}
where $\eta,\mu=\pm1$ denotes the upper/lower branch of the
Keldysh contour.

The tunneling Hamiltonian describing the electron tunneling from
STM tip to the nanotube has a standard form:
\begin{equation}
      H_\mathrm{T}(t)=\sum\limits_{r\alpha\sigma\epsilon}
      \Gamma^{(\epsilon)}(t)\bigl[\Psi_{r\alpha\sigma}^\dagger(0,t)
      c_\sigma(t)
      \bigr]^{(\epsilon)}.
      \label{ham}
\end{equation}
The voltage is taken into account via a time dependence
of the tunneling amplitude (Peierls substitution)
$\Gamma(t)=\Gamma\,e^{i\omega_0 t}$ where
$\omega_0=eV/\hbar$ is the voltage frequency. The superscript
$(\epsilon)$ leaves either operator unchanged $\epsilon=+$, or
transforms it into its Hermitian conjugate $\epsilon=-$.

The total electric current in the nanotube and FL leads can be
expressed through the bosonic field $\phi_{c+}$:
\begin{equation}
      \hat I(x,t)=2ev_{\scriptscriptstyle\rm F}\,\frac{
      \partial_x\phi_{c+}(x,t)}{\sqrt\pi}.
\label{current_operator}
\end{equation}
In Ref. \onlinecite{crepieux}, the nanotube current was computed in terms 
of the bosonic Green's functions, and was found to have the expected
non-linear behavior with voltage. The calculation of this 
current was indeed necessary 
there, because the diagnosis of anomalous charges in the infinite nanotube
required to compute two ratios: the ratio between the auto-correlation 
and the nanotube current, as well as the ratio between the cross-correlation
and the nanotube current. In the present finite frequency scheme which is 
used to extract anomalous charges, the current in both branches of the nanotube
-- which is constant because of the stationary bias, and which equals 
half of the tunneling current in a symmetric setup -- is simply not 
needed. Below, we thus compute the noise auto-correlation and the noise 
cross-correlations.   

\section{Nanotube noise}

\subsection{Unsymmetrized correlator}

The main quantity of our interest will be the nanotube
current-current correlator $S_{xx^\prime}(t,t^\prime)= \langle
\hat I(x,t)\hat I(x^\prime,t^\prime)\rangle$ and using the Keldysh
formalism this quantity can be written as
\begin{equation}
      S_{xx^\prime}(t,t^\prime)\!=\!\left\langle\! T_\mathrm{K}
      \Bigl\{\hat I(x,t^-)\hat I(x^\prime,t^{\prime+})
      e^{-i\int_\mathrm{K} H_\mathrm{T}(\tau)d\tau}\Bigr\}
      \right\rangle\!.
\end{equation}
%
%%%%%%%%%%%%%%%%%%%%%%%%%%%%%%%%%%%%%%%%%%%%%%%%%%%%%%%%%%%%%%%%%%%%%%%%%%%%
Note that here we have purposely chosen the unsymmetrized noise. 
The motivation is as follows. 
When one considers the measurement of the current auto- and cross-correlations
one has to be careful to extract the quantity which is measurable in experiments, 
for this correlator. The ``textbook recipe'' \cite{landau_lifshitz} is to take the
symmetrized current-current correlator: this insures that 
the measured quantity is real. This is indeed the convention which is chosen 
in many reviews about noise \cite{blanter_buttiker}.  

For simplicity, let us forget here 
the spatial dependence of the correlator by considering $x=x'$.
Considers the spectral power of current 
fluctuations (the Fourier transform):
\begin{equation}
S(\omega)=\int S(t)\,e^{i\omega t}\,dt~,
\label{our_convention}
\end{equation}
symmetrized correlations can be expressed as: 
\begin{equation}
S_{sym}(\omega)\equiv {1\over 2}(S(\omega)+S(-\omega))~ .
\end{equation}
However, since the second half of the nineties, the point of view 
-- about which noise correlator is more appropriate -- has shifted. 
When taking a closer insight on this problem one is faced with the fact
that the answer  strictly depends 
on the measurement apparatus which is 
used to measure the noise. 

In Ref. \onlinecite{lesovik_97} a LC circuit was connected inductively
to the mesoscopic device which emits noise, playing the role of a detector. 
There, it was shown that for a passive detector (i.e a detector which is 
in the ground state, which is such that it cannot provide any excitations 
to a measured system) the really measurable quantity is the spectral 
power of current auto--correlations at either positive or negative 
frequencies, depending on the correlator which is considered:
$\langle \hat I(0)\hat I(t)\rangle$ or 
$\langle \hat I(t)\hat I(0)\rangle$.
This results implies that the LC contour (detector) can adsorb
the energy from the measured current but can not excite the measured system
(the later processes correspond to a spectral power at negative
frequencies for the correlator 
$\langle \hat I(0)\hat I(t)\rangle$).

This point of view has been further emphasized 
in two recent theoretical works.
The first of these proposals considers a noisy mesoscopic circuit
which is capacitively coupled to a double dot system 
\cite{aguado_kouvenhoven}, where information on the noise 
is extracted from the measurement inelastic current 
in the double dot, via the transimpedance of the two 
circuits.
The second \cite{gavish} considers a general system 
composed of an antenna (emitter of noise) and a detector, in the 
context of linear response theory. 
Incidentally, both Refs. \onlinecite{aguado_kouvenhoven} and  \onlinecite{gavish}
appeared after the noise review \cite{blanter_buttiker}.
Since these theoretical works, there have been 
contributions \cite{deblock,schoelkopf2}
which point out that in the conditions mentioned in 
\cite{lesovik_97} (i.e. a passive detector), it is the 
non symmetrized contribution  which is experimentally 
measured. 

Ref. \cite{deblock} used a 
superconductor-insulator-superconductor junction SIS
capacitively coupled to the mesoscopic circuit which 
emits noise. We emphasize this particular work because it
chooses the same convention (as in the present work) for the 
the noise correlator $\langle \hat I(t)\hat I(0)\rangle$, 
whose Fourier transform is only relevant at negative 
frequencies.  

\subsection{Zero point fluctuations} 

We are interested in measuring the current-current fluctuations 
in the nanotube due to a tunneling current from the STM tip 
into the nanotube. However, it should be emphasized that the
spectral noise power has finite frequency contributions even if the tip 
is totally decoupled from the nanotube. This point has been 
noticed in a recent paper \cite{trauzettel} where the 
two-terminal noise is computed. There it is explicitly shown that 
although these equilibrium fluctuations (zero point fluctuations 
in our case) vanish at zero frequency, 
there is a finite contribution for the spectral power of the 
symmetrized noise. In thus section we compute 
these correlations and address their relevance
for our non-symmetrized correlator. 

In the absence of coupling to the tip ($\Gamma=0$), we use the expression 
of the current operator Eq. (\ref{current_operator}):  
\begin{equation}
S^0(x-x',t-t') = \partial_x \partial_x' G^{\phi\phi}_{-+}(x,x',t,t')    
\end{equation}
In this equation, the Green's function corresponds to the field $\phi_{c+}$.
In principle, one needs the full expression for the Green's function 
in the presence of the leads. As a starter, let us consider an
infinite nanotube without leads. The Green's function has been 
computed in Ref. \cite{crepieux}:
\begin{equation}
G^{\phi\phi}_{-+}(x,t)= - \frac{1}{8\pi K_{c+}}\sum_r \ln \left( 1 + iv_F{t\over a} + i r K_{c+}{x\over a}\right) 
\end{equation}
In this case,
\begin{equation}
S^0(x,t)= -  {K_{c+}\over 8\pi a^2}\left[ 
\left(v_F{t\over a}+K_{c+}{x\over a}-i\right)^{-2} + \left(v_F{t\over a}-K_{c+}{x\over a}-i\right)^{-2} \right] 
\end{equation}
Consider for simplicity $x=0$.
Taking the Fourier transform, we use Cauchy's theorem and recognize that we pick up a pole 
in the upper half plane when frequencies are positive. We thus conclude that for 
zero or negative frequencies ($\omega\le 0$) there is no contribution to this noise correlator. 
In a similar manner, we find that the correlator $\langle \hat I(0)\hat I(t)\rangle$
has non-zero contributions only for negative frequencies.
Note that the above argument can be generalized for the Green's function of a nanotube 
with leads, which are analyzed in detail in the rest of this paper. 

We thus have a complete agreement with the claims of Ref. \cite{trauzettel}: outside $\omega=0$, 
there is a finite contribution for the symmetrized noise in the absence of the STM tip, 
or, by the same token, in the absence of an applied voltage between 
the tip and the nanotube. 

However, we turn back to the discussion of Ref. \cite{lesovik_97}, and emphasize
that for the correlator $\langle \hat I(t)\hat I(0)\rangle$, only negative frequencies
have a physical meaning. From the discussion above, we found that this correlator 
is zero at negative frequencies. As a result, there is strictly no need to consider 
the effect of zero point fluctuations in the present problem. 
This is a consequence of our choice of setting the temperature to zero 
in our problem. Note that 
this discussion could have been avoided from the start by saying 
that we are considering the contributions to the noise which constitute 
deviations with respect to the zero point result - the excess noise.   

\subsection{Excess noise calculation}

To get the lowest non trivial contribution to the nanotube current
noise, one should expand the exponent in perturbation series in the
tunneling amplitude up to the second order in $\Gamma$. Then
%
%\begin{eqnarray}
%      &&S_{xx^\prime}(t,t^\prime)=-\frac{\Gamma^2}2
%      \sum\limits_{\eta_1\eta_2}\eta_1\eta_2
%      \sum\limits_{r\alpha\sigma\epsilon}
%      \int dt_1dt_2\,e^{-i\epsilon\omega_0(t_1-t_2)}
%      \nonumber\\
%      &&\times
%      \left\langle T_\mathrm{K}
%      \Bigl\{c_\sigma^{(-\epsilon)}(t_1^{\eta_1})
%      c_\sigma^{(\epsilon)}(t_2^{\eta_2})\Bigr\}\right\rangle
%      \\
%      &&\times
%      \left\langle T_\mathrm{K}\Bigl\{
%      \hat I(x,t^-)\hat I(x^\prime,t^{\prime+})
%      \Psi_{r\alpha\sigma}^{(\epsilon)}(0,t_1^{\eta_1})
%      \Psi_{r\alpha\sigma}^{(-\epsilon)}(0,t_2^{\eta_2})
%      \Bigr\}\right\rangle\nonumber.
%\end{eqnarray}
%
expressing the electrons operators in terms of bosonic fields, one
finds that current fluctuations depends only on the time difference
$t-t^\prime$ because of time translational invariance:
$S_{xx^\prime}(t,t^\prime)=S_{xx^\prime}(t-t^\prime)$.
The current fluctuations in the nanotube become:
\begin{eqnarray}
      &&S_{xx^\prime}(t)\!=\!-\frac{e^2v_{\scriptscriptstyle\rm F}^2
      \Gamma^2}{2(\pi a)^2}\sum\limits_{\eta_1\eta_2 r\sigma}
      \!\eta_1\eta_2\!\int \!dt_1dt_2\,
      A_{\eta_1\eta_2}^{r\sigma}(t_1\!-\!t_2)
      \nonumber\\
      &&\times\bigl[ B_{-,\eta_1}^{r\sigma}(x,0,t-t_1)-
      B_{-,\eta_2}^{r\sigma}(x,0,t-t_2)\bigr]
      \nonumber\\
      &&\times\bigl[ B_{+,\eta_1}^{r\sigma}(x^\prime,0,-t_1)-
      B_{+,\eta_2}^{r\sigma}(x^\prime,0,-t_2)\bigr].
      \label{noise_time}
\end{eqnarray}
A real time correlator associated with the tunneling event
at $x=0$ has been introduced:
\begin{eqnarray}
      &&A_{\eta\mu}^{r\sigma}(t)\!=\cos\omega_0t\,
      e^{g_{\eta\mu}^{\sigma}(t)}
      \exp\Bigl[\frac\pi4\sum\limits_{j\delta}
      \tilde G^{\phi\phi}_{j\delta,\eta\mu}(0,0,t)
      \nonumber\\
      &&+
      r\tilde G^{\phi\theta}_{j\delta,\eta\mu}(0,0,t)
      \!+r\tilde G^{\theta\phi}_{j\delta,\eta\mu}(0,0,t)
      \!+\tilde G^{\theta\theta}_{j\delta,\eta\mu}(0,0,t)
      \Bigr]~,\label{factors_time_a}
\end{eqnarray}
together with a correlator associated with propagation of
excitations along the nanotube:
\begin{equation}
      B_{\eta\mu}^{r\sigma}(x,0,t)=
      \partial_x\Bigl[\tilde G^{\phi\phi}_{c+,\eta\mu}(x,0,t)
      +r\tilde
      G^{\phi\theta}_{c+,\eta\mu}(x,0,t)\Bigr].
      \label{factors_time_b}
\end{equation}
Here $\tilde G^{\phi\phi}_{j\delta,\eta\mu}
(x,x^\prime,t-t^\prime)=\bigl\langle T_\mathrm{K}\bigl\{
\phi_{j\delta}(x,t^\eta)\phi_{j\delta}(x^\prime,
t^{\prime\mu})\bigr\}\bigr\rangle -\frac12\langle
\phi^2(x,t)\rangle-\frac12\langle\phi^2(x^\prime,t^\prime)\rangle$
is the Keldysh Green's function for the nanotube bosonic field
$\phi_{j\delta}(x,t)$ (similar definitions hold for other
combinations of bosonic fields), $g^\sigma_{\eta\mu}(t)$ is the
Keldysh Green's function for STM tip bosonic field defined in
Eq.~(\ref{STM_gr}).

We now define the spectral power of current fluctuations as
$S_{xx^\prime}(\omega)=\int S_{xx^\prime}(t)\,e^{i\omega t}\,dt$, 
i.e. including the dependence on spatial coordinates.
Taking the Fourier transform of Eq.~(\ref{noise_time}), which is a
convolution, can be done assuming that the Fourier transform of
$A_{\eta\mu}^{r\sigma}(t)$ and $B_{\eta\mu}^{r\sigma}(x,0,t)$
are known. One therefore gets:
\begin{eqnarray}
      &&S_{xx^\prime}(\omega)=-\frac{e^2v_{\scriptscriptstyle\rm F}^2
      \Gamma^2}{2(\pi a)^2}\sum\limits_{\eta_1\eta_2r\sigma}
      \eta_1\eta_2
      \nonumber \\
      &&\>\>\times\Bigl(\tilde A_{\eta_1\eta_2}^{r\sigma}(0)
      \tilde B_{-,\eta_1}^{r\sigma}(x,0,\omega)
      \tilde B_{+,\eta_1}^{r\sigma}(x^\prime,0,-\omega)
      \nonumber\\
      &&\>\>\>-\,\tilde A_{\eta_1\eta_2}^{r\sigma}(-\omega)
      \tilde B_{-,\eta_2}^{r\sigma}(x,0,\omega)
      \tilde B_{+,\eta_1}^{r\sigma}(x^\prime,0,-\omega)
      \nonumber\\
      &&\>\>\>-\,\tilde A_{\eta_1\eta_2}^{r\sigma}(\omega)
      \tilde B_{-,\eta_1}^{r\sigma}(x,0,\omega)
      \tilde B_{+,\eta_2}^{r\sigma}(x^\prime,0,-\omega)
      \nonumber\\
      &&\>\>\>+\,\tilde A_{\eta_1\eta_2}^{r\sigma}(0)
      \tilde B_{-,\eta_2}^{r\sigma}(x,0,\omega)
      \tilde B_{+,\eta_2}^{r\sigma}(x^\prime,0,-\omega)
      \Bigr),
      \label{noise_fr}
\end{eqnarray}
where $\tilde A(\omega)$, $\tilde B(x,0,\omega)$ are Fourier
transforms, $\tilde A(\omega)=\int A(t)e^{i\omega t}\,dt$ and
similarly for $\tilde B(x,0,\omega)$, of factors $A(t)$ and $B(x,0,t)$
defined in Eqs.~(\ref{factors_time_a}) and (\ref{factors_time_b}).

\section{Green's functions}

Let us now concentrate on the calculation of the bosonic Green's
functions in the nanotube. Going from the bosonic
Hamiltonian of Eq.~(\ref{b_ham}) to the Lagrangian in imaginary time
$\tau=it+\delta$ which is in general dependent on both fields
$\phi_{j\delta}$ and  $\theta_{j\delta}$ \cite{crepieux}.
To derive the Green's functions for  the two fields
$\phi_{j\delta}(x,\tau)$ ($\theta_{j\delta}(x,\tau)$)
it is convenient to integrate the partition function over
the field $\theta_{j\delta}(x,\tau)$ ($\phi_{j\delta}(x,\tau)$).
The corresponding actions follow:
\begin{equation}
      S_{\phi_{j\delta}}=\frac12\int dxd\tau\,K_{j\delta}\Bigl[
      v_{j\delta}^{-1}\,(\partial_\tau\phi_{j\delta})^2
      \!+v_{j\delta}(\partial_x\phi_{j\delta})^2\Bigr],
      \label{action_ph}
\end{equation}
\begin{equation}
      S_{\theta_{j\delta}}=\frac12\int dxd\tau K_{j\delta}^{-1}\Bigl[
      v_{j\delta}^{-1}(\partial_\tau\theta_{j\delta})^2+
      v_{j\delta}(\partial_x\theta_{j\delta})^2
      \Bigr],
        \label{action_th}
\end{equation}
where $K_{j\delta}(x)$ and $v_{j\delta}(x)$ are functions of
the coordinate $x$. From these actions, one can conclude that the
Green's function, $G^{\phi\phi}_{j\delta}(x,x^\prime,\bar \omega)$
for the field $\phi_{j\delta}(x,\tau)$
($G^{\phi\phi}_{j\delta}(x,x^\prime,\bar \omega)$ is the Fourier
transform of $G^{\phi\phi}_{j\delta}(x,x^\prime,\tau)$ in
imaginary time), obeys the following differential equation:
\begin{equation}
      \Bigl[\frac{K_{j\delta}(x)}{v_{j\delta}(x)}\,\bar
      \omega^2\!-\!
      \partial_x v_{j\delta}(x)K_{j\delta}(x)\partial_x\Bigr]
      G^{\phi\phi}_{j\delta}(x,x^\prime,\bar \omega)\!=
      \!\delta(x\!-\!x^\prime).
      \label{gr_eq}
\end{equation}

A similar equation holds for
$G^{\theta\theta}_{j\delta}(x,x^\prime,\bar \omega)$, making the
transformation $K_{j\delta}(x)\rightarrow K_{j\delta}^{-1}(x)$ as
is obvious from Eqs.~(\ref{action_ph}) and (\ref{action_th}). It
follows from Eq.~(\ref{gr_eq}) that the Green's function
$G^{\phi\phi}_{j\delta}(x,x^\prime,\bar \omega)$ is a continuous
function and its derivative $\partial_x
G^{\phi\phi}_{j\delta}(x,x^\prime,\bar \omega)$ has a jump at
$x=x^\prime$:
\begin{equation}
      -v_{j\delta}(x)K_{j\delta}(x)\partial_x
      G^{\phi\phi}_{j\delta}(x,x^\prime,\bar \omega)
      \Bigr|_{x=x^\prime-\epsilon}^{x=x^\prime+\epsilon}=1.
      \label{jump}
\end{equation}
To solve Eq.~(\ref{gr_eq}) with the condition given by Eq.~(\ref{jump}),
one chooses a model where the interaction parameter is a step like
function: $K_{j\delta}(x)=K_{j\delta}^{\scriptscriptstyle\rm
N}$ in the nanotube ($|x|<\frac L2$) and
$K_{j\delta}(x)=K_{j\delta}^{\scriptscriptstyle\rm L}$ in the
leads ($|x|>\frac L2$). Then looking for solution for
$|x^\prime|<\frac L2$ in the form:
\begin{eqnarray}
      G^{\phi\phi}(x,x^\prime,\bar \omega)\!=\!
      \left\{
      \begin{array}{ll}
      Ae^{\frac{|\bar \omega|}{v_{\scriptscriptstyle\rm L}}x},&
      x<-\frac{L}{2},
      \\
      Be^{\frac{|\bar \omega|}{v_{\scriptscriptstyle\rm N}}x}+\!
      Ce^{-\frac{|\bar \omega|}{v_{\scriptscriptstyle\rm N}}x},&
      -\frac{L}{2}<x<x^\prime,
      \\
      De^{\frac{|\bar \omega|}{v_{\scriptscriptstyle\rm N}}x}+\!
      Ee^{-\frac{|\bar \omega|}{v_{\scriptscriptstyle\rm N}}x},&
      x^\prime<x<\frac{L}{2},
      \\
      Fe^{-\frac{|\bar \omega|}{v_{\scriptscriptstyle\rm L}}x},&
      x>\frac{L}{2},
      \end{array}
      \right.
\label{ABCDEF}
\end{eqnarray}
where the Green's function $G^{\phi\phi}(x,x^\prime,\bar \omega)$
vanishes at $x\rightarrow\pm\infty$, corresponding to outgoing
boundary conditions (we have omitted subindex $j\delta$ to shorten
the notations). Solving all matching conditions at $x=\pm
\frac{L}2$ and $x=x^\prime$, one specifies all coefficients
$A,B,...$ in Eq. (\ref{ABCDEF}). In order to obtain the crossed
Green's functions $G^{\phi\theta}(x,x^\prime,\bar\omega)$ and
$G^{\theta\phi}(x,x^\prime,\bar\omega)$, one uses the equation
of motion for the field $\phi(x,\tau)$ and $\theta(x,\tau)$
generated from Hamiltonian of Eq.~(\ref{ham}). This leads to:
\begin{eqnarray}
      &&G^{\phi\theta}(x,x^\prime\bar\omega)\!=-
      \frac1{\bar\omega}\frac{v(x)}{K(x)}
      \,\partial_x G^{\theta\theta}(x,x^\prime,\bar\omega),
      \\
      &&G^{\theta\phi}(x,x^\prime\bar\omega)\!=\!-
      \frac{v(x)K(x)}{\bar\omega}
      \,\partial_x G^{\phi\phi}(x,x^\prime,\bar\omega).
\end{eqnarray}
Performing the inverse Fourier transform
one obtains the bosonic Green's functions in real time. From the
real time Green's functions the Keldysh Green's
functions matrix elements can be specified. Its two time arguments $t$ and $0$ are
assigned to the lower/upper ($+/-$) branch of the Keldysh contour. This procedure has been
described in detail in Ref. \onlinecite{crepieux}, and shall therefore
not be repeated.
In order to find the noise correlation
in the nanotube one needs only the real time Green's functions at the
origin for factor $A_{K}^{r\sigma}(t)$ of Eq.~(\ref{factors_time_a})
and one needs the Fourier transform of
the coordinate derivative of the Green's functions in the leads for
factors $B_K^{r\sigma}(x,x^\prime,t)$ of Eq.~(\ref{factors_time_b}). Using the scheme of
calculation depicted above one finds:
\begin{equation}
      \tilde G_{j\delta}^{\phi\theta}(0,0,t)=
      \tilde G_{j\delta}^{\theta\phi}(0,0,t)=0,
\end{equation}
and
\begin{eqnarray}
      &&\tilde G^{\phi\phi}_{j\delta}(0,0,t)=-\frac1{2\pi
      K_{j\delta}^{\scriptscriptstyle\rm N}}\Biggl\{
      \ln\Bigl(1+\frac{iv_{\scriptscriptstyle\rm F}t}a
      \Bigr)\nonumber
      \\
      &&+\sum\limits_{r=\pm1}\sum\limits_{n=1}^\infty
      b_{j\delta}^n\ln\Bigl[1+\frac{iv_{\scriptscriptstyle\rm
      F}t}{a+irnK_{j\delta}^{\scriptscriptstyle\rm N}L}
      \Bigr]\Biggr\},
      \label{gr_origin}
\end{eqnarray}
where $b_{j\delta}=(K_{j\delta}^{\scriptscriptstyle\rm
N}-K_{j\delta}^{\scriptscriptstyle\rm
L})/(K_{j\delta}^{\scriptscriptstyle\rm
N}+K_{j\delta}^{\scriptscriptstyle\rm L})$. The Green's function
$\tilde G^{\theta\theta}_{j\delta}(0,0,t)$ can be obtained by
transformation $K_{j\delta}^{\scriptscriptstyle\rm N}\rightarrow
(K_{j\delta}^{\scriptscriptstyle\rm N})^{-1}$ of the total factor
of Eq.~(\ref{gr_origin}). Another necessary Green's functions has
the form:
\begin{eqnarray}
      &&\partial_xG_{++}^{\phi\phi}(x,0,\omega)\!=\!
      \frac{i\,sgn(x)}{(K^{\scriptscriptstyle\rm N}\!
      +\!K^{\scriptscriptstyle\rm L})v_{\scriptscriptstyle\rm L}}\,
      f(|\omega|)
      e^{i\frac{|\omega|}{v_{\scriptscriptstyle\rm
      L}}(|x|-\frac{L}2)},\nonumber
      \\
      &&\partial_xG_{-+}^{\phi\phi}(x,0,\omega)\!=\!
      \frac{i\,sgn(x)\Theta(\omega)}{(K^{\scriptscriptstyle\rm N}\!
      +\!K^{\scriptscriptstyle\rm L})v_{\scriptscriptstyle\rm L}}
      \Bigl(f(\omega)e^{i\frac{\omega}{v_{\scriptscriptstyle\rm
      L}}(|x|-\frac{L}2)}\!-\!C.c.\Bigr),\nonumber
      \\
      &&\partial_xG_{++}^{\phi\theta}(x,0,\omega)\!=\!
      \frac{-i\,sgn(\omega)K^{\scriptscriptstyle\rm N}}{
      (K^{\scriptscriptstyle\rm N}\!
      +\!K^{\scriptscriptstyle\rm L})v_{\scriptscriptstyle\rm L}}\,
      f(|\omega|)
      e^{i\frac{|\omega|}{v_{\scriptscriptstyle\rm
      L}}(|x|-\frac{L}2)},\nonumber
      \\
      &&\partial_xG_{-+}^{\phi\theta}(x,0,\omega)\!=\!
      \frac{-i\Theta(\omega)K^{\scriptscriptstyle\rm N}}{
      (K^{\scriptscriptstyle\rm N}\!
      +\!K^{\scriptscriptstyle\rm L})v_{\scriptscriptstyle\rm L}}
      \Bigl(\!f(\omega)e^{i\frac{\omega}{v_{\scriptscriptstyle\rm
      L}}(|x|-\frac{L}2)}\!+\!C.c.\!\Bigr),\nonumber
      \\
      &&\qquad\qquad \>\>f(\omega)=
      \frac{e^{i\frac{\omega}{v_{\scriptscriptstyle\rm N}}\frac{L}{2}}}
      {1-b\,e^{i\frac{\omega}{v_{\scriptscriptstyle\rm N}}L}}\,,
\end{eqnarray}
where $\pm$ refers to the branch of the Keldysh contour and $\Theta(\omega)$ is the Heaviside function.
To ease notations, the index $j\delta$ has been omitted from
each coefficient $K^{{\scriptscriptstyle\rm
N}({\scriptscriptstyle\rm L})}$ and $v_{{\scriptscriptstyle\rm
N}({\scriptscriptstyle\rm L})}$.
The remaining set of Keldysh Green's functions can be found from
symmetry properties:
\begin{eqnarray}
      &&\partial_xG^{\phi\phi}_{--}(x,0,\omega)=
      [\partial_xG^{\phi\phi}_{++}(x,0,-\omega)]^*,
      \\
      &&\partial_xG^{\phi\phi}_{+-}(x,0,\omega)=
      [\partial_xG^{\phi\phi}_{-+}(x,0,-\omega)]^*,
\end{eqnarray}
with the same relations for $\partial_x
G^{\phi\theta}_K(x,0,\omega)$ Green's functions.

\section{Noise correlations for an infinite nanotube}

Let us first consider the case of infinite nanotube $L\rightarrow
\infty$ where the noise correlations are measured in the Luttinger
liquid. In this case, in all factors $\tilde
A_{\eta\mu}^{r\sigma}(\omega)$ and $\tilde
B_{\eta\mu}^{r\sigma}(x,x^\prime,\omega)$ calculated in previous
sections one should insert interaction parameters to be equal
$K_{j\delta}^{\scriptscriptstyle\rm
N}=K_{j\delta}^{\scriptscriptstyle\rm L}=K_{j\delta}$.
This corresponds to an homogeneous nanotube Hamiltonian~(\ref{b_ham})
with  constant velocity $v_{j\delta}$, without leads.
Then, substituting $\tilde A_{\eta\mu}^{r\sigma}(\omega)$, $\tilde
B_{\eta\mu}^{r\sigma}(x,x^\prime,\omega)$ into
Eq.~(\ref{noise_fr}) one gets the following expression for the
noise:
\begin{eqnarray}
      &&S_{xx^\prime}(\omega)=\frac{e^2\Gamma^2}{(\pi a)^2}\,
      \frac{K_{c+}^2+sgn(x) sgn(x^\prime)}2
      \nonumber\\
      &&\>\>
      \times\Theta(-\omega)\,e^{i\omega\tau_-}
      \Bigl(\tilde A_{+-}(|\omega|)+\tilde A_{-+}(-|\omega|)\Bigr),
\label{noise_correlations_infinite}
\end{eqnarray}
where $\tau_-=(|x|-|x^\prime|)/v_{c+}$ is defined to be the retardation time.

According to Ref.~\onlinecite{lesovik_97}, only negative
frequencies are entering in the expression for the spectral noise
correlator $S_{xx^\prime}(\omega)$ at zero temperature.
Physically, this corresponds to the measurable part of our
definition $S(\omega)=\int \langle \hat I(t)\hat
I(0)\rangle\,e^{i\omega t}\,dt$ for the auto-correlation function.

As is explicit in Eq. (\ref{noise_correlations_infinite}), 
the current cross-correlator is a complex quantity, which
reflects the fact that the product of two Hermitian operators in
general is not a Hermitian operator itself. Yet the noise cross-correlator 
is a quantity which should in principle be measured 
experimentally. This is a similar issue which was encountered
for the autocorrelation noise: what is the quantity which is 
physically measurable in experiments ? 
In general this correlator is a complex quantity for non symmetric 
positions $x\neq -x'$ so it cannot
be directly measurable quantity even 
at positive frequencies.

A general theory of noise cross-correlation measurements goes beyond 
the scope of this paper. Here we only cite existing results.  
When considering the same model 
as in Ref. \cite{lesovik_97}, for the cross correlator at two arbitrary 
positions $x$ and $x'$, with an LC circuit inductively coupled to our 
mesoscopic circuit at these two locations, one finds that  \cite{lebedev}
the noise correlator should be symmetrized with respect to the 
two positions where current are being measured:
\begin{equation}
S^{meas}_{xx^\prime}(\omega)={1\over 2}[S_{xx^\prime}(\omega)+S_{x^\prime x}(\omega)]
\end{equation} 
Note that these issues can be avoided if one considers the two positions
where currents are being measured to be symmetrically located with 
respect to the injection point of the STM ($x=-x'$).

\begin{figure}[h]
\psfrag{sx}{$~~~~S_{x,x}$}
\psfrag{w}{$-\omega/\omega_0$}
\epsfxsize 7 cm
\centerline{\epsffile{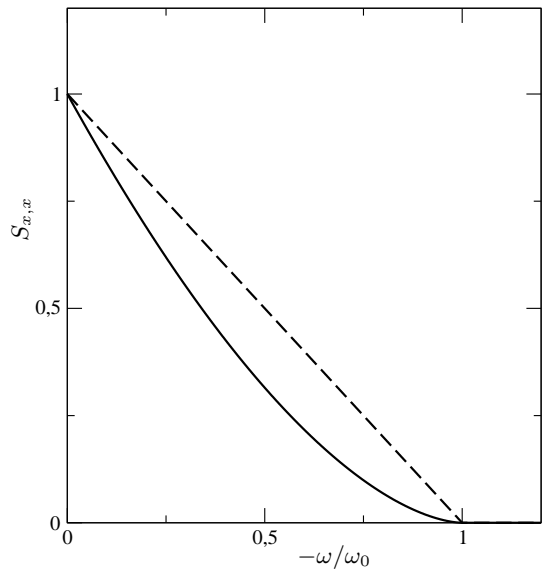}}
\caption{Finite frequency noise auto-correlation for an infinite nanotube,
in the non interacting case ($K_{c+}=1$ dashed lines), as well as
in the case of repulsive interactions  ($K_{c+}=1/3$ full line).
The noise is normalized to the zero frequency value. In the case 
of $K_{c+}=1/3$, the cross-correlations $-S_{x,-x}$ have the same 
dependence on $\omega$.
\label{fig4}}.
\end{figure}

In the plots of Figs. \ref{fig4} and \ref{fig5} of this section
and the following section the trivial phase $e^{i\omega\tau_-}$ will thus
be omitted. Note that it is equal to $1$ for a measurement geometry where 
$|x|=|x^\prime|$ (equal length between the two current measurements 
on either side of the nanotube and the tip). 

The factors $\tilde
A_{+-}(\omega)$ and $\tilde A_{-+}(\omega)$ entering
expression (\ref{noise_correlations_infinite}) are defined as:
\begin{equation}
      \tilde A_{+-}(\omega)=\int\limits_{-\infty}^\infty
      \frac{\cos\omega_0t\,e^{i\omega t}}{
      (1-iv_{\scriptscriptstyle F}t/a)^{\nu+1}}\;dt,
\label{a+-}
\end{equation}
and $\tilde A_{-+}(\omega)=\tilde A_{+-}^*(-\omega)$, where
\begin{equation}
      \nu=\sum\limits_{j\delta}\frac18\Bigl(
      K_{j\delta}+\frac1{K_{j\delta}}\Bigr).
\end{equation}
In the limit of a vanishing cutoff $a\rightarrow 0$, one can compute this integral
explicitly and find the current fluctuations in the form:
\begin{eqnarray}
      &&S_{xx^\prime}(\omega)=\frac{K_{c+}^2+sgn(x)sgn(x^\prime)}2
      \;e^{i\omega\tau_-}
      \nonumber\\
      &&\times\frac{2e^2\Gamma^2}{\pi v_{\scriptscriptstyle\rm F}^2}
      \left(\frac{a}{v_{\scriptscriptstyle\rm
      F}}\right)^{\nu-\!1}\!\Theta(|\omega_0|-|\omega|)\,
      \frac{(|\omega_0|-|\omega|)^\nu}{{\bf\Gamma}(\nu+1)}.
\label{noise_correlations_infinite_final}
\end{eqnarray}
According to Eq. (\ref{noise_correlations_infinite_final}), the noise cross correlator
and the noise auto-correlations differ by a prefactor only. The convention for measuring
the currents has been chosen as in Ref. \onlinecite{crepieux}. The positive direction is the
same for $x>0$ and for $x<0$. Consequently the noise autocorrelation is positive while the
noise cross-correlations are negative. Recall that in Ref. \onlinecite{crepieux}, it was stated that
the usual convention for measuring noise correlations, i.e. in Hanbury-Brown type experiments
\cite{hbt}, is to chose the positive direction  to correspond
to both currents flowing away from the point of injection\cite{martin_landauer_buttiker}.
This means, for instance that when considering $S_{x,-x}(\omega=0)<0$, quasi-particles are moving
away from the injection point and give rise to positive Hanbury-Brown and Twiss correlations,
as for photons.
\vskip1.5cm
Consider a nanotube with all interaction parameters $K_{j\delta}=1$. This corresponds
to the non-interacting case. Physically, this situation is reached when, for instance,
the nanotube is put in close proximity with a metallic gate or with a heavily doped
semiconductor substrate. According to Eq. (\ref{noise_correlations_infinite_final})
there is an autocorrelation signal corresponding to electrons being emitted either
right or left of the emitter. However, there is no cross-correlation signal whatsoever,
even at finite frequencies.  In order to measure a cross-correlation signal,
it would be necessary to go to higher order (fourth order) in the tunneling amplitude
\cite{martin_landauer_buttiker}, where scattering theory predicts negative Hanbury-Brown
and Twiss correlations, a result in agreement with experimental observations\cite{henny}.

When considering the auto-correlation at finite frequencies, one recovers
(Fig. \ref{fig4}) the known
finite frequency spectrum of noise associated with non-interacting
mesoscopic systems \cite{yang}: the noise decreases linearly with frequency, until
the characteristic frequency $eV/\hbar$ is reached. The derivative of the noise
bears a singularity at this point, which has been diagnosed experimentally
in a diffusive mesoscopic conductor \cite{schoelkopf}.

Next, one chooses $K_{c+}<1$. All other interactions parameters are equal to 
$1$: for the spin sector, this assumes that time reversal symmetry applies.
For convenience, in the plots of Fig. \ref{fig4}, we have omitted the prefactor
$(K_{c+}^2+sgn(x)sgn(x^\prime))/2$ in order to plot the auto-correlation and
the cross-correlation noise spectrum in a single plot.
At $\omega=0$, one recovers the result of Ref. \onlinecite{crepieux}: positive
Hanbury-Brown and Twiss correlations due to entangled quasi-particles flowing at
opposite ends of the nanotube. At frequencies larger than $\omega_0$, the noise
is equal to zero as in the non-interacting case. 

For $\omega<\omega_0$, the noise differs
from the non-interacting case. 
First of all, recall that in {\it chiral} Luttinger liquids as in 
the fractional quantum Hall effect\cite{chamon_freed_wen}, 
the finite frequency spectrum associated with quasiparticle tunneling 
gives rise to a singularity at  $\omega=\nu eV/\hbar$ ($\nu$ is the filling factor)
due to the tunneling density of states of Laughlin quasiparticles.
Recall also that, when tunneling occurs between two quantum Hall fluids, 
the noise at $\omega<\nu eV/\hbar$ has a power law behavior, 
because of the tunneling density of states of electrons. 
The situation is quite similar when electrons are injected into a non-chiral 
Luttinger liquid: the spectrum for $\omega<\omega_0$ (Fig. \ref{fig4}) has 
a power law behavior with no singularity. The spectral density of noise is thus 
always lower than that of the non-interacting case.

From the point of view of the auto-correlation noise, there is therefore
no qualitative difference between the spectrum for electron tunneling 
into a chiral and a non-chiral Luttinger liquid. 
From the point of view of the noise cross-correlation spectrum,
one concludes that positive correlations persist at finite frequencies.
In what follows, one should enquire whether such features remain robust in
the presence of FL leads.

\section{Noise correlations for a nanotube connected to leads}

Consider now the case where the nanotube has a finite length
($L\gg a$), and  where it is connected to a FL leads with
$K_{j\delta}^{\scriptscriptstyle\rm L}=1$. Keeping the same conventions as
in  the preceding sections, the expression for the spectral density of noise
-- measured in the leads -- reduces to:
\begin{eqnarray}
      &&S_{xx^\prime}(\omega)=
      \,\frac{K_{c+}^2|\phi_+(\omega)|^2\!+\!
      sgn(x)sgn(x^\prime)|\phi_-(\omega)|^2}{(1+K_{c+})^2}
      \nonumber\\
      &&\times\frac{2e^2\Gamma^2}{(\pi a)^2}\;
      \Theta(-\omega)\,e^{i\omega\tau_-
      }\Bigl(\tilde A_{+-}(|\omega|)+\tilde
      A_{-+}(-|\omega|)\Bigr),
\end{eqnarray}
where $\phi_\pm(\omega)=(1\pm (K_{c+}-1)/(K_{c+}+1) e^{i\omega
L/v_{c+}})^{-1}$ reflects the presence of the leads.
%%%%%%%%%%%%%%%%%%%%%%%%%%%%%%%%%%%%%%%%%%%%%%%%%%%%%%%%%%%%%%%%%

In principle, the functions $\tilde A_{+-}(\omega)$ also contain
information on the presence of the leads. In what follows, we will neglect 
the effect of the leads only in  $\tilde A_{+-}(\omega)$ under specific 
assumptions. 
Consider the expression of A(t) in Eq. (\ref{factors_time_a}), which reflect
the local tunneling contribution in the noise correlator.
It can be decomposed as a product 
of three functions, when one considers the expression for the Green functions in 
the presence of the leads: a) 
the $\cos (\omega_0 t)$ prefactor;
b) a function which is the Green's function in the absence
of leads; this function only has a dependence on t 
and the cutoff $v_F/a$;
c) a function which has all the information about the leads with  
a time scale $\tau_\mathrm{L}= v_{c+}/L$.
Our approximation is to forget about this last term   
because when the Fourier transform is taken, if the voltage
frequency $\omega_0$ is large, the variation of the factor c) will 
not matter: the combination of the oscillation and the slowly 
decaying part of the (infinite nanotube) contribution b) will
provide the dominant contribution. Note that  
this approximation is justified if the Fourier frequency which is sampling 
$\tilde A_{+-}(\omega)$ is smaller than (or at least less than) $\omega_0$.      
The validity limit of this approximation is 
that $\tau_\mathrm{V}$ is small compared to $\tau_\mathrm{L}$. 

Here, one shall use the expression for $\tilde A_{+-}(\omega)$ of Eq. (\ref{a+-})
corresponding to an infinite nanotube. Finally, one gets the following
expression for current fluctuations:
\begin{eqnarray}
      &&S_{xx^\prime}(\omega)\!=\!\frac{e^2\Gamma^2}{
      \pi v_{\scriptscriptstyle\rm F}^2}\!
      \left(\frac{a}{v_{\scriptscriptstyle\rm F}}\right)^{\nu\!-\!1}
      \!\!e^{i\omega\tau_-}
      \Theta(|\omega_0|\!-\!|\omega|)
      \frac{(|\omega_0|\!-\!|\omega|)^\nu}{{\bf \Gamma}(\nu\!+\!1)}
      \nonumber\\
      &&\times\!\left(\!\frac1{1\!-\!(1\!-\!K_{c+}^{-2})\!\sin^2\omega\tau_\mathrm{L}}
      \!+\!\frac{sgn(x)sgn(x^\prime)}{1\!-\!(1\!-\!K_{c+}^{2})
      \!\sin^2\omega\tau_\mathrm{L}}\!\right),
\label{noise_leads_final}
\end{eqnarray}
where $\tau_\mathrm{L}=L/2v_{c+}$ is the traveling time needed for
Luttinger liquid excitations to reach the leads.

\begin{figure}[h]
\psfrag{S}{$-S_{x,-x}$}
\psfrag{w}{$-\omega/\omega_0$}
\psfrag{a}{$a)$}
\psfrag{b}{$b)$}
\epsfxsize 7 cm
\centerline{\epsffile{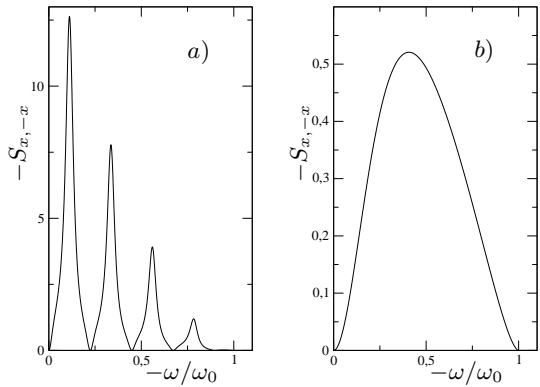}}
\caption{Finite frequency noise correlations for a nanotube connected to leads.
a) case where $\tau_\mathrm{L}=14\tau_\mathrm{V}$; b)case where $\tau_\mathrm{L}=\tau_\mathrm{V}$. The noise is normalized
according to the frequency independent prefactor of Eq.~(\ref{noise_leads_final}).
\label{fig5}}
\end{figure}

These results are illustrated in Fig. \ref{fig5}, for different parameters.
In addition to the frequency scale imposed by the voltage bias, when considering
the inhomogeneous Luttinger model for the FL leads, there appears a
new frequency scale $\tau_\mathrm{L}^{-1}$ associated with the time of flight defined
above. The first observation which can be made from the two plots of Figs.
\ref{fig5}a  and \ref{fig5}b is that at zero frequency, the noise
cross-correlation vanish. The work of Ref. \onlinecite{crepieux} had expressed
suspicion that this may be the case, although
a full computation of noise correlations was not provided there. The issue of the
presence of FL leads for the two-terminal conductance of a Luttinger
liquid leads was addressed several years ago \cite{safi_maslov}.

At finite frequencies however, the noise correlations are always positive
(with the ``usual'' convention for current signs), or zero.
Positive noise correlations are reminiscent of systems where the two constituents
(here pairs of quasi-particles)
of a particle (here an electron) are separated into two different branches.
Positive correlations where encountered previously in branched
normal-superconducting junctions, when the two constituent electrons
of a Cooper pair are split into two normal metal terminals \cite{torres_martin}.

The fact that the spectral power vanishes at zero frequency seems
rather natural when using a second order approximation scheme in the tunneling
amplitude $\Gamma$. Indeed, the second order calculation takes into
account coherent transport associated with {\it only one}
electron at a time injected into nanotube. It is a Poissonian result: 
there is no correlation between two successive electrons.  
At the same time, the zero
frequency fluctuations are directly related to the cross-correlations
of the {\it total} charge transmitted to the right FL
lead and to the left FL lead. Let us inject an electron into the 
nanotube at time $t=0$.
The cross correlations of the total charge transmitted to the FL
leads $\hat Q_\mathrm{L}(t)=\int_0^t\hat I(x<0,t^\prime)
dt^\prime$ and $\hat Q_\mathrm{R}(t)=\int_0^t\hat I(x>0,t^\prime)
dt^\prime$ are defined as:
\begin{equation}
      \langle \hat Q_\mathrm{L}(t)\hat Q_\mathrm{R}(t)\rangle=
      \int S_{xx^\prime}(\omega)\,\frac{\sin^2(\omega
      t/2)}{(\omega/2)^2}\,\frac{d\omega}{2\pi}.
\end{equation}
In the limit $t\rightarrow\infty$, one has $\langle \hat
Q_\mathrm{L}(t)\hat Q_\mathrm{R}(t)\rangle=tS_{xx^\prime}(0)$
(since $\sin^2(\omega t/2)/(\omega/2)^2\rightarrow 2\pi
t\delta(\omega)$ in this limit). Since one has injected only one
electron, it can be transmitted as whole either to the left or to
the right FL lead and thus $\langle \hat Q_\mathrm{L}(t)\hat
Q_\mathrm{R}(t)\rangle=0$. This simple observation accounts for
the vanishing of the current fluctuations at zero frequency. 
Of course the next order in perturbation theory takes into account
the coherence effect between two consecutively injected electrons, 
which may result in finite current fluctuations at zero frequency.
This is for instance true for electron injection in a FL, where the 
cross-correlations at $\omega=0$ are negative.

The next observation coming from Eq.~(\ref{noise_leads_final}) is
that there are two different limits for electron transport
depending on the ratio of the voltage time
$\tau_\mathrm{V}=\hbar/eV$ and the transport time $\tau_\mathrm{L}$.
The situation for the case where $\tau_\mathrm{L}\gg\tau_\mathrm{V}$ is
depicted on Fig. \ref{fig5}a where a sequence of resonances at
$\omega\tau_\mathrm{L}=(2p+1)\pi/2$ appears in the current
fluctuations. These resonances can be accounted by the analogy
with multiple Andreev reflection on the boundaries between the 
nanotube and the FL leads \cite{ines_annales}.

The injection of an
electron, say at $t=0$, results in the creation of two anomalously charged
excitations propagating in opposite directions toward the boundaries. 
The characteristic width of the anomalously charged quasiparticle 
wave packets, $\delta x_{Q_\pm}$ is specified by the voltage time: 
$\delta x=v_{c+}\tau_\mathrm{V}$. Recall that these quasiparticle wave 
packets consist of collective electron-hole excitations.  
In the limit $\tau_\mathrm{L}\gg \tau_\mathrm{V}$, the width of these
anomalously charged wave packets is much smaller then the length of
the nanotube ($L\gg\delta x$). As a result, it takes a finite time
$\tau_\mathrm{L}$ for these excitations to reach the boundary. At
the boundary, these excitations can either be transmitted to the FL leads
or can be reflected back to the nanotube, resulting in the splitting of these
anomalously charged wave packets. The transmitted part (which consists of
multiple electron-hole excitations, now in the FL lead) is then measured,
resulting in the first peak of Fig. \ref{fig5}a. On the other hand, 
the reflected part then moves to the opposite boundary and reaches it after a time
$t=3\tau_\mathrm{L}$. There it is again transmitted or reflected from this
boundary. The transmitted parts in the FL leads on both side give rise to 
another signal (second maximum in Fig. \ref{fig5}a).  
Such multiple reflection processes therefore result in the sequence of
peaks in the spectral power curve at the corresponding frequencies. 
The amplitude of the peaks are lower, the higher the number of reflections
inside the nanotube region, as expected. In all of the above, the 
cross-correlations signal is always positive in the Hanbury-Brown 
and Twiss sense.  

In the opposite limit $\tau_\mathrm{L}\ll
\tau_{\mathrm{V}}$ (see Fig.~\ref{fig5}b), the size of the anomalously charged wave packets
is much larger than the length of the nanotube and there is no room for 
resonances to occur in the cross-correlation spectral power curve.
In the following one can argue that the positive noise correlation 
signal in either situations -- many resonances or one resonance -- 
can be used to identify anomalous charges without any ambiguity. 
However, note that given our assumptions on the computation of 
$\tilde A_{+-}(\omega)$, results are to be trusted mostly for multiple 
bounces: in the case of a single bounce, a correction of the 
density of states due to the presence of the leads 
should in principle be included. 
%%%%%%%%%%%%%%%%%%%%%%%%%%%%%%%%%%%%%%%%%%%%%%%%%%%%%%%%%%%%%%%%%%%%%%%%%%%%%%%%%%
\vskip1cm

\section{Conclusion}

The purpose of the present paper has been to further analyze the effect of electron injection
in a one-dimensional correlated electron system, chosen to be a metallic carbon nanotube.
A crucial question was to ask whether the positive noise correlations survive at zero
frequencies, when leads are included in the model. The negative answer to this question
motivated us to analyze the full finite frequency spectrum of auto-correlation and
cross-correlation noise.

The presence of positive correlations in the finite frequency
cross-correlations is a manifestation of non-local behavior in an interacting one-dimensional
system, which was stressed in Ref. \onlinecite{crepieux} in the case of vanishing frequencies.
Here, we have shown that a finite frequency analysis allows to recover such information.
In addition, in is particularly interesting to uncover the role of the different time scales
which govern the dynamics of electron injection in a Luttinger liquid connected to leads.
In one extreme, one obtains a simple vanishing of zero frequency noise at the
cutoff frequency, while in the other extreme, the noise spectrum
shows a diffraction pattern due to the interference of quasi-particles traveling
in the central region.

In all of the above, we conclude that the finite frequency analysis of noise allows to clearly
identify the anomalous charges which are associated with the chiral modes traveling
in this non-chiral Luttinger liquid. In an experiment, these could be detected by choosing a
frequency where the noise autocorrelation and the noise cross-correlations
give a maximum as a function of frequency. A suggestion for identifying the interaction parameter
-- and thus the anomalous charges -- is to tune the frequency so  that $\omega \tau_\mathrm{L}= (2p+1)\pi/2$
($p$ integer), so that the sine function which enters in Eq. (\ref{noise_leads_final}) is equal
to one. Then, one can measure experimentally -- and compute --
the ratio $|S_{x,-x}/ S_{x,x}|= (1-K_{c+}^4)/(1+K_{c+}^4)$. to extract the chiral
charges $Q_\pm =(1\pm K_{c+})/2$. Note that the fact that such chiral charges enter the
Fabry-Perot interpretation \cite{ines_annales} for recombination into electron charges at the Luttinger
liquid interfaces gives a motivation for their identification as elementary charges
of the non-chiral Luttinger liquid.

While this was in progress, we noticed a work\cite{trauzettel} dealing with two terminal measurements
in a Luttinger liquid containing an impurity, connected to FL leads. There is a claim that a charge
$e^*=K_{c+}e$ (the translation to the nanotube interaction parameter in ours) can be identified in a
periodic structure of the noise. It is likely that their results are connected to the present results.
In some sense, the injection of an electrons may play the same role as the presence of an impurity.
However, our approach has the advantage that it merely relies on the comparison of the spectrum
for autocorrelation and for cross-correlation noise. Note that the present results apply just as well
to a one-dimensional quantum wire constructed from semiconductor cleaved-edge heterostructures such as the ones
used to measure spin-charge separation\cite{yacoby}.

The present work, like many others, has described the contacts as one-dimensional leads which are
adiabatically connected to the Luttinger liquid. Extensions of the present work could include
a more general description of the connection between the Fermi liquid leads
and the nanotube. This has been achieved using both a coherent and an incoherent approach for
the fractional quantum Hall effect\cite{chamon_fradkin}. Also, as mentioned in our previous work,
more complications could arise due to screening\cite{screening} of the interactions by the FL leads
or even by the tip. The former should not play any role as long as the size of the nanotube is larger
than the screening length, which should be comparable to the lattice constant or cutoff parameter $a$.
Concerning screening effect due to the tip, however, the fact that a tunneling
geometry is chosen should minimize its effects.
A final remark is that this finite frequency diagnosis could be also simulated by a zero-frequency
noise measurement operated on a nanotube which is subject to the superposition of a DC and an 
AC bias from the tip, as was recently illustrated for the fractional quantum 
Hall effect \cite{ACDC}.    

\acknowledgements
T. M. acknowledges fruitful discussions with Fabrizio Dolcini and Hermann Grabert 
during the workshop on ``Correlation Effects in Nano-wires'' (Venice 2004).
A.L. thanks the Ecole-Normale Landau Institute agreements and CNRS for his stay at the 
Centre de Physique Th\'eorique. He also acknowledges financial support from the Landau Scholarship
of the FZ Julich and from the Russian Dynasty Foundation Fund.

\end{document}